\documentclass[preprint,endfloats*]{revtex4}
\nonstopmode

\usepackage{amsmath}
\usepackage{amsbsy}

\newcommand{\dirint}[3]{\ensuremath{\langle #1|#2|#3\rangle}}

\newcommand{\half}{{\textstyle\frac{1}{2}}}

\newcommand{\bs}{\boldsymbol}

\newcommand{\EX}[1]{\Psi_{#1}}  
\newcommand{\ul}{\underline}

\begin{document}
\title{Is time-dependent density functional theory formally exact?}
\author{J. Schirmer}
\affiliation{Theoretische Chemie, Physikalisch-Chemisches Institut,
Universit\"at Heidelberg, \\
D-69120 Heidelberg, Germany}
\author{A. Dreuw}
\affiliation{Institut f\"ur Physikalische und Theoretische Chemie,
Universit\"at Frankfurt, Germany\\
D-60439 Frankfurt}
\date{\today}

\begin{abstract}
The general expectation that, in principle, the time-dependent density functional theory (TDDFT) be an exact
formulation of the time-evolution of an interacting $N$-electron system 
is critically reexamined. 
It is demonstrated that the previous TDDFT foundation, resting on four theorems by Runge and Gross (RG) 
[Phys.~Rev.~Lett.~52, 997(1984)], is invalid because undefined phase factors corrupt the RG
action integral functionals. Our finding confirms much of  
a previous analysis by van Leeuwen~[Int. J. Mod. Phys. B~15, 1969(2001)]. 

To analyze the RG theorems and other aspects of TDDFT, an utmost simplification of the KS
concept has been introduced, in which the ground-state density is obtained from a single KS equation for one
spatial (spin-less) particle. The time-dependent (td) form of this radical Kohn-Sham (rKS) scheme, 
which has the same validity status as the ordinary KS version, has proved to be a valuable  
tool for analysis. The rKS concept is used to clarify also the alternative
non-variational formulation of td KS theory. 
We argue that it is just a formal theory, allowing one to
reproduce, but not predict the time-development of the exact density of the 
interacting $N$-electron system.

Besides the issue of the formal exactness of TDDFT, it is shown that both the static and time-dependent
KS linear response equations neglect the particle-particle ($p$-$p$) and hole-hole ($h$-$h$) 
matrix elements of the perturbing operator. For a local (multiplicative) operator this does not
lead to a loss of information due to a remarkable general property of local operators. Accordingly, no
logical inconsistency arises with respect to DFT, because DFT requires any external potential to be local.
For a general non-local operator the error resulting from the neglected matrix elements is 
of second order in the electronic repulsion.
\end{abstract}
\maketitle
\newpage

\section{Introduction}

Over the last decade, time-dependent density functional theory (TDDFT) has become 
an extremely popular method to compute electronic excitation energies and response properties
of ever bigger molecules and clusters (see, for example, Refs.~\cite{sun99:480,fur00:1717,dre02:12070}). 
The method and its foundations were already worked out in the nineteen eighties, primarily
in papers by Zangwil and Soven~\cite{zan80:1561}, Runge and Gross~\cite{run84:997},
and Gross and Kohn~\cite{gro85:2850}. More recently, various efficient computer codes have been 
developed~\cite{bau96:454,jam96:5134,str98:8218,gis98:2556,toz98:10180,gor99:2785} and made available as parts
of major quantum chemistry program packages. At present, one witnesses intense activities worldwide 
aiming both at the further development of methodological aspects and the computational efficiency of 
the codes. 

Besides the relatively modest computational expense, a major boost for the 
advancement of the method has been the assurance that TDDFT is a formally exact 
theory~\cite{cas95:155,pet96:1212,hir99:291}, that is, the TDDFT results would become exact 
if the exact time-dependent exchange-correlation (xc) potentials were available. In practice, of course,
one always has to use approximate xc potentials, and therefore one has to be prepared for smaller or 
larger errors in the computational results. 
There is a widely held confidence that any problems encountered
with the TDDFT method are only caused by imperfections of the underlying xc potentials, 
a belief prevailing even as
some more severe problems became apparent, such as in 
the description of Rydberg excitations~\cite{toz98:10180,cas00:8918}, the treatment of extended
$\pi$ systems~\cite{cai02:5543}, the absence
of double (and higher)
excitations~\cite{toz00:2117,mai04:5932}, and the $1/R$ dependence of charge-transfer (CT) excitation 
energies~\cite{toz99:859,sob03:73,dre03:2943,dre04:4007}.
These failures have triggered efforts to modify the 
xc potentials accordingly and thereby remedy the respective problems. Most of this work
has been confined to the so-called adiabatic approximation, in which the time-dependence 
enters the xc functionals only via the time-dependent (td) density functions. But also 
the development of time- or energy-dependent
xc functionals beyond the adiabatic approximation~\cite{mai02:023002,ull04:28,mai04:5932} 
has been envisaged.

On the other hand, TDDFT has never obtained a similarly accepted status of uncontested validity as
the original (time-independent) density functional theory (DFT) developed
by Hohenberg and Kohn (HK)~\cite{hoh64:864} and by Kohn and Sham (KS)~\cite{koh65:1133}.
The foundations of TDDFT, as formulated
by Runge and Gross (RG) in Ref.~\cite{run84:997}, 
have been constructed largely in terms analogous to the HK and KS concepts of DFT. However,  
elusive notions such as td v-representability and non-interacting 
v-representabilty were clearly in need of further mathematical 
clarification~\cite{koh86:1993,lee99:3863,coh05:32515}.    
More recently, the RG foundations of TDDFT were challenged by Rajagopal~\cite{raj96:3916},
van Leeuwen~\cite{lee98:1280}, and Harbola and Banerjee~\cite{har99:5101}, 
after it was realized that the kernel of the xc functional 
in the RG formulation violates causality~\cite{gross:1995,nalewajski:1996}.  
A critical review
of the RG action integral functionals by van Leeuwen~\cite{lee98:1280,lee01:1969}  
revealed basic deficiencies. Presently, an alternative formulation of TDDFT,
being essentially a KS-type approach without implying a variational principle of the HK type,
is viewed as a valid foundation~\cite{gross:1995,lee01:1969,hei02:9624,bur05:062206}. 
 
In this paper we will take a new look at the foundations of TDDFT. For our review
we use a simple analytical device, referred to as radical Kohn-Sham (rKS) formulation, which
is as legitimate as the usual $N$-electron KS theory. Not obscured by intricacies such as  
td v-representability etc.,
the rKS concept allows us to analyze both the static DFT and TDDFT in an utmost transparent way. 
What we find, confirms van Leeuwen's criticism of the RG foundation of TDDFT, but also proves
the non-variational form of TDDFT illusory.   
    
An outline of the paper is as follows. The starting point of our study (Sec.~2) is the observation
that the TDDFT equations, more specifically, the time-dependent Kohn-Sham (KS) linear response (LR)
equations, neglect matrix elements of the perturbing (external) potential of the $h$-$h$ or 
$p$-$p$ type, where $h$ and $p$ refer to occupied (hole) and unocccupied (particle) KS orbitals, respectively. 
Because
the exact linear response depends on all matrix elements, the TDDFT results appear to be deficient
irrespective of the choice of the xc potential. The same situation arises in the case of 
a static (time-independent) perturbation, as is analyzed in Sec.~3. Here the problem would even
challenge the well-founded (time-independent) DFT. The 
resolution of that puzzle in Sec.~3 is a very instructive confirmation of the logical consistency 
of DFT. 
In Sec.~4 we introduce the rKS concept, in which the GS density is not
determined from the density of $N$ non-interacting electrons, but from a single KS equation
for one (spin-less) particle. In Sec.~5 the rKS formulation is used to analyze the RG theorems 
and other aspects of TDDFT. A reader primarely interested in the issue of the validity of TDDFT might skip
Secs.~2 and 3 and leap directly to Sec.~4. A summary of our results and
some conclusions are given in the final Sec.~6.

\section{Comparison of exact and Kohn-Sham linear response}

The TDDFT formalism has been presented in various ways in previous 
work~\cite{gro90:255,cas95:155,jam96:5134,str98:8218,hir99:291}
where the reader is referred to for an overview and further details. Because TDDFT (in linear response form)
is similar to the time-dependent Hartree-Fock (TDHF) or random-phase
approximation (RPA)~\cite{mcl64:844,dal66:291,row68:153}, it is rewarding to consult 
also previous TDHF derivations 
(see, for example, Ring and Schuck~\cite{Ring:1980}). A few basic notions
pertinent to the ensuing discussion will be given in the following.

For an $N$-electron system (atom or molecule) having a non-degenerate ground state
$\left|\Psi_0 \right>$ the (exact) ground state (GS) density matrix, $\bs{\gamma}$, 
is given by 
\begin{equation}
\label{eq:rhomat}
\gamma_{pq} = \dirint{\EX{0}}{c^{\dagger}_qc_p}{\EX{0}}
\end{equation}
Here the second-quantized operators $c^{\dagger}_p(c_p)$ are associated with  
one-particle states (spin-orbitals) $\phi_p$. As a particular choice, we will
consider the KS orbitals arising from the KS one-particle equations associated with
the ground state of the system under consideration,
\begin{equation}
\label{eq:kseq}
h^{KS} \phi_i(\bs{r},s) = \{-\half \nabla^2 + v(\bs{r}) + J[\rho](\bs{r}) 
                   + v_{xc}[\rho](\bs{r})\} \phi_i(\bs{r},s) = \epsilon_i \phi_i(\bs{r},s)
\end{equation}
Here $\bs{r}$ and $s$ denote spatial and spin variables, respectively, $v(\bs{r})$ 
is the one-particle operator for the electron-nuclei interaction, 
$J[\rho](\bs{r})$ is the Coulomb operator, and $v_{xc}[\rho](\bs{r})$ is the KS 
exchange-correlation potential.
The exact GS density function, $\gamma (\bs{r})$, is obtained from the density matrix
elements (\ref{eq:rhomat}) according to 
\begin{equation}
\label{eq:rhoexact}
\gamma (\bs{r}) = \sum_{p,q} \sum_s \gamma_{qp}\, \phi^*_p(\bs{r},s)\phi_q(\bs{r},s)
\end{equation}
By contrast to the exact density matrix, the KS density matrix, $\bs{\rho}$, is derived from the
KS determinant
\begin{equation}
\label{eq:ksdet}
\left|\Phi_0^{KS} \right> = \left| \phi_1 \dots \phi_N \right|
\end{equation}
according to 
\begin{equation}
\label{eq:ksmat}
\rho_{pq} = \dirint{\Phi_0^{KS}}{c^{\dagger}_qc_p}{\Phi_0^{KS}}
\end{equation}
In the KS orbital representation assumed here the KS density matrix assumes the simple 
diagonal form
\begin{equation}
\label{eq:ksdiag}
\rho_{pq} = \delta_{pq} n_p
\end{equation}
where $n_p = 0,1$ denote KS occupation numbers. 
The KS density function,
\begin{equation}
\label{eq:ksdensfun}
\rho (\bs{r})  =  \sum_{p,q} \sum_s \rho _{qp}\, \phi^*_p(\bs{r},s)\phi_q(\bs{r},s)\\
                      =  \sum_k \sum_s \left | \phi_k(\bs{r},s) \right |^2 n_k
\end{equation}
is devised to reproduce the exact density function, that is, $\gamma (\bs{r}) \equiv \rho (\bs{r})$,
provided the correct exchange-correlation potential is used in Eq.~(\ref{eq:kseq}).
Whereas, at least in principle, the exact and KS density functions are identical, the density
matrices necessarily must differ. As is well recognized (see, for example, \cite{cas95:155,hir99:291}), 
the two entities differ with respect to a basic property: the KS density matrix, deriving from a 
single determinantal wave function, is idempotent, that is, $\bs{\rho}^2 = \bs{\rho}$,
whereas the exact density matrix is not, $ \bs{\gamma}^2 \neq \bs{\gamma}$. 

To discuss the linear response (LR) theory let us consider an additional time-dependent (td) 
external (``driving'')
potential of the form
\begin{equation}
\label{eq:extpot}
\hat{u} = \hat{d}\,f(t) 
\end{equation}
where $\hat{d} = d(\bs{r})$ is a local (multiplicative) operator and $f(t)$ 
is a scalar time dependent function 
(vanishing for $t<0$).
In the notation of second quantization, the corresponding $N$-electron operator, $\hat{D} =\sum_i^N \hat{d}(i)$,
can be written as
\begin{equation}
\label{eq:dop}
\hat{D} = \sum_{r,s} d_{rs} c^{\dagger}_rc_s
\end{equation}
where $d_{rs} =  \dirint{\phi_r}{\hat{d}}{\phi_s }$ denote the one-particle matrix elements of $\hat{d}$.
Now let us distinguish particle-hole ($p$-$h$) and 
$h$-$p$ matrix elements, $d_{ak}, d_{ka}$, from $p$-$p$ and $h$-$h$ elements, $d_{ab}, d_{kl}$. 
Here and in the following we use the notation in which the
subscripts $a,b,c,\dots$ and $i,j,k,\dots$ denote unoccupied (virtual) and occupied KS orbitals, 
respectively, while the subscripts $p,q,r,\dots$ refer to the general case. 
As will be discussed below, the KS linear response contribution to the density depends only 
on the $p$-$h$ (and $h$-$p$) matrix
elements of the driving potential, whereas the exact linear response contribution is a linear 
function of all matrix elements. 

Let us first inspect the exact case.
Upon Fourier transformation the linear response of the exact density matrix can be written as
(see, for example,~\cite{Fetter:1971})
\begin{equation}
\label{eq:exresp}
\delta \gamma _{pq}(\omega) = \sum_{n \neq 0} \frac{\dirint{\EX{0}}{c^{\dagger}_qc_p}{\EX{n}}
                                     \dirint{\EX{n}}{\hat{D}}{\EX{0}}}
                                    {\omega - E_n + E_0 +i\eta} - 
                             \frac{\dirint{\EX{0}}{\hat{D}}{\EX{n}}
                                     \dirint{\EX{n}}{c^{\dagger}_qc_p}{\EX{0}}}
                                    {\omega + E_n - E_0 +i\eta}
\end{equation}
Here $\left| \EX{n} \right>$ and $E_n$ denote excited energy eigenstates and eigenvalues of the original (undisturbed)
Hamiltonian $\hat{H}$; the complex infinitesimal $i \eta$ is required for the definiteness of the Fourier 
transforms between the time and energy domain. For the special operator $\hat{D}$ considered here, 
the transition moments appearing in the numerators on the
right-hand side of Eq.~(\ref{eq:exresp}) take on the form
\begin{equation}
\label{eq:amplitudes}
\dirint{\EX{n}}{\hat{D}}{\EX{0}} = \sum_{p,q} d_{pq} \dirint{\EX{n}}{c^{\dagger}_pc_q}{\EX{0}} 
\end{equation}
and it is obvious that the $p$-$p$ and $h$-$h$ contributions, $\dirint{\EX{n}}{c^{\dagger}_ac_b}{\EX{0}}$ and 
$\dirint{\EX{n}}{c^{\dagger}_kc_l}{\EX{0}}$, need not vanish. Using 
many-body perturbation theory (MBPT) for $\left | \Psi_0 \right>$ and $\left | \Psi_n \right>$ 
based on the familiar M\o ller-Plesset decomposition of $\hat{H}$
(and adopting for a moment Hartree-Fock (HF) one-particle states)
one may readily establish that non-vanishing contributions appear for the first time in second order. 
For example, one finds
\begin{equation}
\label{eq:orders}
\dirint{\EX{n}}{c^{\dagger}_ac_b}{\EX{0}} = O(2)
\end{equation}
for single excitations deriving from the HF configurations $c^{\dagger}_a c_j \left | \Phi^{HF}_0 \right >$.
An analogous result is found in the case of the $h$-$h$ amplitudes. Moreover, the exact response 
comprises contributions arising from double (and higher) excitations, the corresponding 
$p$-$p$ and $h$-$h$ amplitudes being here even of first order.

Now let us turn to the KS response theory. In the formulation given by Gross and Kohn~\cite{gro85:2850} 
(adopting here a slightly deviating notation) the linear 
response to the KS density function is given by
\begin{equation}
\label{eq:gk1}
\delta \rho (\bs{r},\omega) = \int \chi (\bs{r},\bs{r}';\omega) v
                                  ^{eff}_1(\bs{r}',\omega) d\bs{r}'
\end{equation}
Here 
\begin{equation}
\label{eq:gk2}
\chi (\bs{r},\bs{r}';\omega) = \sum_{pq} \sum_{s,s'} (n_p -n_q) 
                            \frac{\phi^{*}_p(\bs{r},s) \phi_q (\bs{r},s)\phi^{*}_q(\bs{r}',s')
                              \phi_p(\bs{r}',s')}
                             {\omega -\epsilon_q + \epsilon_p +i\delta}
\end{equation}
is referred to as the KS density-density response function and
\begin{equation}
\label{eq:gk3}
v^{eff}_1(\bs{r},\omega) = u(\bs{r},\omega) + J[\delta \rho](\bs{r}) + \delta v_{xc}(\bs{r},\omega)
\end{equation}
is the first-order effective potential comprising
the (Fourier transformed) external perturbation of Eq.~(\ref{eq:extpot}),
the (first-order) change of the Coulomb potential, $J[\delta \rho](\bs{r})$, 
and of the xc potential, $\delta v_{xc}(\bs{r},\omega)$, the latter two contributions being linear expressions  
in $\delta \rho(\bs{r},\omega)$.

Inserting the first part of $v^{eff}_1(\bs{r},\omega)$, 
that is, the ``driving'' potential, $u(\bs{r},\omega)$, in the rhs of Eq.~(\ref{eq:gk1}) yields 
\begin{eqnarray}
\label{eq:gk4}
\int \chi (\bs{r},\bs{r}';\omega) u(\bs{r}',\omega) d\bs{r}'
                           =  \sum_{a,k} \sum_s \left( \frac{\phi^{*}_k(\bs{r},s) \phi_a (\bs{r},s)}
                            {\omega - \epsilon_a + \epsilon_k +i\delta }\, d_{ak} \right.\\
\nonumber                    \left.      - \frac{\phi^{*}_a(\bs{r},s) \phi_k (\bs{r},s)}
                            { \omega + \epsilon_a - \epsilon_k +i\delta}\, d_{ka} \right) f(\omega)
\end{eqnarray}
Obviously, here the  $p$-$p$ and $h$-$h$ matrix elements of $\hat{d}$ have been projected out and only
$p$-$h$ (and $h$-$p$) matrix elements, $d_{ak}$, enter (as inhomogeneities) the linear KS response
equations.  As we have seen,
the exact linear response to the density function, $\delta \gamma (\bs{r},\omega)$, which may be
written in a form analogous to Eq.~(\ref{eq:gk1}),
\begin{equation}
\label{eq:gk6}
\delta \gamma (\bs{r},\omega) = \int \overline{\chi}(\bs{r},\bs{r}';\omega) u(\bs{r}',\omega) d\bs{r}'
\end{equation} 
is a linear function of all matrix elements, $d_{pq}$, 
of the perturbing potential.                
Here the exact density-density response function~\cite{gro85:2850}, $\overline{\chi}(\bs{r},\bs{r}';\omega)$, 
is related to $\delta \bs{\gamma}$ of Eq.~(\ref{eq:exresp}) according to 
\begin{equation}
\label{eq:gk7}
\int \overline{\chi}(\bs{r},\bs{r}';\omega) d(\bs{r}') d\bs{r}'= \sum_{p,q} \sum_s \delta \gamma_{pq}(\omega)
                                             \phi_q^{*}(\bs{r},s)\phi_p(\bs{r},s)
\end{equation}
This result evokes the question if equating $\delta \rho (\bs{r},\omega)$ and 
$\delta \gamma(\bs{r},\omega)$
is permitted at all.
It seems that in the KS linear response equations
the information associated with the $p$-$p$ and $h$-$h$ matrix elements of the external 
(driving) potential is lost and, thus, these equations have to be viewed as an approximation 
even in the case of an exact exchange correlation potential. 

Let us inspect the situation in
the more general and transparent matrix formulation of TDDFT 
(see, for example, \cite{str98:8218,hir99:291}). 
Here the KS response equations are written in the form of a matrix commutator relation,
\begin{equation}
\label{eq:matresp}
 \omega\, \delta \bs{\rho} = [\bs{h}, \delta \bs{\rho}] + [\delta \bs{h},\bs{\rho}]
                                        + [\bs{d}, \bs{\rho}] 
\end{equation}
where $\bs{h}$ and $\bs{d}$ denote the matrix representations of the (unperturbed) KS Hamiltonian and 
the perturbing potential (time-independent part), respectively, and $\delta \bs{h}$ is the change of the 
KS Hamiltonian linear in $\delta \rho$. Note that due to the form of $\bs{\rho}$ the commutator
$[\bs{d}, \bs{\rho}]$ on the rhs of Eq.~(\ref{eq:matresp}) projects out the $p$-$p$ and $h$-$h$ matrix elements
of $\bs{d}$. 
Arranging the 
$p$-$h$ and $h$-$p$ matrix elements of $\delta \bs{\rho}$ and of $\hat{d}$ in columns (vectors), 
\begin{equation}
\label{eq:matnot}
\delta \ul{\rho} = \left( \begin{array}{c}
                                  \delta \ul{\rho}_{ph}\\
                                  \delta \ul{\rho}_{hp}
                                  \end{array}
                            \right), \,\,\,\,
\ul{d} = \left( \begin{array}{c}
                                  \ul{d}_{ph}\\
                                  \ul{d}_{hp}
                                  \end{array}
                            \right)
\end{equation}
the linear response equation for $\delta \ul{\rho}$ takes on the familiar RPA 
form,
\begin{equation}
\label{eq:mateqs}
\left( \begin{array}{cc}
                     \omega - \bs{A} & -\bs{B}\\
                     -\bs{B}^*       & -\omega - \bs{A}^*
         \end{array} 
         \right) \delta \ul{\rho} =  \ul{d} 
\end{equation}
The elements of the matrices $\bs{A}$ and $\bs{B}$, being  related to 
the functional derivatives of the Coulomb and xc potentials 
of the KS Hamiltonian, have been specified elsewhere (see, for example, Ref.~\cite{bau96:454}, 
and Sec. 3). The
information on the perturbing external potential enters the response equations (\ref{eq:mateqs}) only via 
the vector $\ul{d}$. Thus, it is manifest
that only the $p$-$h$ and $h$-$p$ matrix elements of the perturbing potential 
come into play. Does the neglect of the $h$-$h$ and $p$-$p$ matrix elements of the perturbing 
operator mean that 
TDDFT is not formally exact?  The same problem occurs 
in the case of a time-independent (static) perturbation, and here it
would even challenge the logical consistency of DFT itself, more specifically, the universality of 
the HK energy functionals. In the next Sec. 3 we will consider the static case and see how
the apparent contradiction to the universality of the DFT functionals can be resolved.

\section{Time-independent Kohn-Sham response theory}
In this section we consider the problem of the loss of the $h$-$h$ and $p$-$p$ 
matrix elements in
the simpler static case of  a (small) time-independent external perturbation, $\hat{u}$.

The linear KS response equations  are obtained here as a special case ($\omega = 0$)
of the more general
time-dependent equations (\ref{eq:matresp}):
\begin{equation}
\label{eq:cpks}
[\bs{h}, \delta \bs{\rho}] + [\delta \bs{h}, \bs{\rho}] + 
[\bs{u},\bs{\rho}]  
= \bs{0}
\end{equation}
Likewise, these equations can be deduced via first-order perturbation theory for the ground state KS
orbital, 
also referred to as coupled-perturbed 
Kohn-Sham (CPKS) theory (see Casida\cite{cas95:155} and references therein). 
As above $\bs{h}$ and $\bs{u}$ denote the matrix representations of the unperturbed KS Hamitonian and
the perturbing potential, respectively. 
The KS density matrix, $\bs{\rho}$,
associated with the unperturbed ground state is diagonal, $\rho _{pq}=\delta_{pq} n_p$;
$\delta \bs{\rho}$ denotes the first-order change
in the KS density matrix. Finally,  
$\delta \bs{h}$  is the matrix representation of the
linear change of the KS Hamiltonian, 
\begin{equation}
\label{eq:dhks}
\delta h = J[\delta \rho](\bs{r}) + \delta v_{xc}(\bs{r})
\end{equation}
A basic assumption of the KS linear response theory is that the xc part of $\delta h$
can be expanded according to
\begin{equation}
\label{eq:dhks2}
\delta v_{xc} = v_{xc}[\rho + \delta \rho] - v_{xc}[\rho] 
              = \int \frac{\delta v_{xc}[\rho](\bs{r})}{\delta \rho (\bs{r'})} \delta \rho (\bs{r'}) d\bs{r'}
                  + O(\delta \rho^2) 
\end{equation}
in terms of $\delta \rho$ (and, possibly, gradients, $\nabla \delta \rho$, and higher derivatives).
Here $\delta \rho$ is  related to the first-order density matrix,
$ \delta \bs{\rho}$,
according to
\begin{equation}
\label{eq:deltarho}
\delta \rho(\bs{r}) = \sum_{a,k} \sum_s
                      \left \{ \phi^*_k (\bs{r},s) \phi_a (\bs{r},s) \delta \rho_{ak} + 
                       \phi^*_a (\bs{r},s) \phi_k (\bs{r},s) \delta \rho_{ka} \right\}
\end{equation}

Proceeding in the usual way, one now may
evaluate the $p$-$h$ and $h$-$p$ matrix elements of $\delta h$,
\begin{equation}
\label{eq:deltahmat1}
(\delta \bs{h})_{pq} = \int \delta h (\bs{r}) \phi^*_p(\bs{r}) \phi_q(\bs{r})\,d \bs{r}
\end{equation}
which leads to linear expressions in the density matrix elements
$\delta \rho_{rs}$:
\begin{equation}
\label{eq:deltahmat2}
(\delta \bs{h})_{pq} = \sum_{rs} M_{pq,rs}\, \delta \rho_{rs} 
\end{equation}
Here the index pairs, (pq) or (rs), are either of $p$-$h$ or $h$-$p$ type.
Finally,
introducing $\delta \bs{h}$ in that form in Eq.~(\ref{eq:cpks}) one arrives at the 
desired KS linear response equations, reading in matrix form analogous to
Eqs.~(\ref{eq:matnot},\ref{eq:mateqs}),
\begin{equation}
\label{eq:CPKSmateqs}
\left( \begin{array}{cc}
                      \bs{A}   &  \bs{B}\\
                      \bs{B}^* & \bs{A}^*
         \end{array} 
         \right) \ul{\delta} \rho =  - \ul{u} 
\end{equation}
Here, the matrix elements of $\bs{A}$ and $\bs{B}$ are given by
\begin{equation}
\label{eq:CPmatel}
A_{ak,bl} = (\epsilon_a - \epsilon_k) \delta_{ab} \delta_{kl} + M_{ak,bl}, \,\, B_{ak,lb} = M_{ak,lb}
\end{equation}
As in Eq.~(\ref{eq:matnot}), the $p$-$h$ and $h$-$p$ matrix elements of
$\delta \rho$ and $u$ are arranged to form column vectors,
\begin{equation}
\ul{\delta} \rho = \left( \begin{array}{c}
                                  \ul{\delta}\rho_{ph}\\
                                  \ul{\delta}\rho_{hp}
                                  \end{array}
                            \right), \,\,\,\,
\ul{u} = \left( \begin{array}{c}
                                  \ul{u}_{ph}\\
                                  \ul{u}_{hp}
                                  \end{array}
                            \right)
\end{equation} 
As in the td case, the perturbation enters the set of linear equations via the 
inhomogeneity vector, $\ul{u}$, in which the $p$-$p$ and $h$-$h$ matrix elements of $u$ are absent.

It appears that we are facing a paradox here: on the one hand, we have just applied a valid first-order 
perturbation theory to the density within the KS framework arriving at 
a seemingly deficient result; on the other hand, the full solution of the KS eigenvalue problem for the 
perturbed Hamiltonian must yield 
the exact density, so that also the result of first-order perturbation theory cannot be incorrect.
The answer to this puzzle is that no information on the perturbing potential is lost in the 
absent $p$-$p$ and $h$-$h$ matrix elements provided that 
the perturbation is a local (multiplicative) operator \cite{Head-Gordon:04}. This is due to a remarkable, 
though apparently not widely known  
property of local operators which may be stated as follows:\\
\emph{Theorem 1}.
A local operator, $v=v(\bs{r})$,
is uniquely determined up to a constant by its $p$-$h$ (and $h$-$p$) matrix elements with 
respect to a complete one-particle basis and an arbitrary partitioning of that basis into
occupied (hole) and unoccupied (particle) one-particle states.\\ 
A simple proof of this assertion is given in the 
Appendix. An interesting aspect here is that the proof assures merely the uniqueness (up to a 
constant) of the local operator but does not offer a way to reproduce the operator from
its $p$-$h$ matrix elements. It seems that for such a reconstruction one needs one
of the diagonal blocks, that is, either $h$-$h$ or $p$-$p$, in addition to the $p$-$h$ block. 
Thus, the logical status of theorem 1 resembles that of the 
HK and KS theorems which prove the existence of universal xc functionals
but do not provide for means to construct the functionals.

Theorem 1 assures that the loss of the $p$-$p$ and $h$-$h$ matrix elements in the CPKS 
equations is no contradiction to the formal exactness of the theory, provided that the
external potentials are local (multiplicative). Clearly, this observation applies also to the td 
KS linear response considered in Sec. 2. The restriction to local external (one-particle) potentials 
is a basic and well understood consistency requirement of DFT. The universality of the kinetic
energy and xc functionals hinges on the condition that the external potential functionals are of the form
\begin{equation}
\label{eq:expotfun}
E_{v}[\rho] = \int  v(\bs{r}) \rho(\bs{r}) d\bs{r}
\end{equation}

It should be recalled that quantum theory is essentially non-local, and 
many physically important interactions are not of the local type. For example,
the interaction of electrons with an electromagnetic field involves the 
momentum operators, $\ul{p}_j = -i \nabla _j$. It is common practice, to apply
the usual CPKS and TDDFT methods also for non-local external potentials (see, for example,
\cite{buh99:91,nee01:11080,tun03:11024}). In that case one should be aware that the loss of the $p$-$p$ and $h$-$h$ matrix elements
of the external operator introduces indeed an error beyond the
approximation for the functional, which is of second order in the Coulomb repulsion. In the
td KS linear response this error affects only the transition moments. In particular, it
destroys the equivalence between the so-called length and velocity forms of the 
transition moments, because the former is associated with a local operator and the latter
with a non-local operator.

In principle, DFT can be extended to account for non-local external potentials as well. For
this purpose the non-local potential must be incorporated ``a priori'' into the
HK and KS formalism, that is, the HK and KS functionals have to be defined from the 
outset for the $N$-electron
system under the action of the external potential. 
This would lead to modified functionals being now
specific to the considered non-local potential. In other words, the functionals would    
depend on the respective non-local external potentials. An important example for the 
necessity to deal with non-local operators is the presence of magnetic fields. As a
systematic approach referred to as current density functional theory (CDFT) 
one here considers functionals that depend not only
on the density but also on the current density \cite{gho88:1149,vig90:235,col94:271}. 
As another possibilty of dealing with non-local external potentials, Gilbert~[\cite{gil75:2111}] 
and Levy~[\cite{lev79:6062}] have considered density-matrix dependent functionals.

Let us  briefly inspect how the CPKS equations will change if the   
the xc potential depends directly on a non-local perturbing potential, $u$.
Obviously, this would lead to an additional contribution to $\delta h$ of the 
form 
\begin{equation}
\label{eq:deltahext}
\delta \tilde{v}_{xc} = w[\rho,u](\bs{r})
\end{equation}
and, thus, to another inhomogeneity term in the linear response equation (\ref{eq:CPKSmateqs}). 
Here $\rho$ is the unperturbed GS density. 
Because the additional inhomogeneity contribution depends on $u$, the full information
on $u$ can be restored, reconciling 
the (first-order) result of the KS linear response with the exact result. 

As a more general aspect, the non-local potential problem shows that
the CPKS equations, while justified as a valid first-order perturbation theory for the
KS approach to determine the ground state density of the system plus perturbation, may not
be seen as physical response equations for the interacting $N$-electron 
system (in that case they should apply also to non-local
perturbations). This admonishes us to be wary of the prospect that the td KS equations
can describe the time evolution of the system in response to a td perturbation.

\section{A radical Kohn-Sham version}

The KS formulation is a clever way to transform the problem of finding the density minimizing
the HK functional into the determination of the ground state of an associated non-interacting
N-particle system. While providing a good basis for
practical computational schemes, the usual KS formulation still does not achieve its full  
theoretical potential. 
In fact, one may proceed to a radical KS approach, in which the mapping of the
exact density is not onto that of a non-interacting N-particle system but rather to the density of 
a single particle.
Whereas such a radical KS formulation will be less suitable as a starting point
for the approximate treatment of the exact ground state density, it may serve as a valuable
analytical tool to clarify various aspects of DFT and, in particular, TDDFT. It should be noted
that the idea of such an obvious extension of the usual KS approach is not new, though apparently 
little known. Already in 1984 it was used by Levy \emph{et al.}~\cite{lev84:2745} to discuss
asymptotic properties of the xc potential.  
  
In the usual (N-particle) KS formulation the kinetic energy contribution, $T[\rho]$, to the HK functional is 
substituted by the kinetic energy functional
\begin{equation}
\label{eq:KSKE}
T_{S}[\rho] = \sum_{i,s}  \int \psi^*_i(\bs{r},s)(-\half \nabla^2)\psi_i(\bs{r},s)\, d\bs{r}
\end{equation}
of a non-interacting $N$-particle system, the density being obtained according to
\begin{equation}
\label{eq:rhoKS}
\rho (\bs{r}) = \sum_{k} \sum_s  \psi^*_k(\bs{r},s) \psi_k(\bs{r},s)
\end{equation}
as the density function associated with the Slater determinant $\left|\Phi \right> = 
\left|\psi_1 \psi_2 \dots \psi_N \right| $ of orthonormal orbitals $\psi_i,  i=1, \dots, N$.
As is well-known, the functional (\ref{eq:KSKE}) can 
be made unambiguous using the Levy constrained search (LCS) definition \cite{lev79:6062} (see also 
Parr and Yang \cite{Parr:1989}):
\begin{equation}
\label{eq:LevyCS}
T_{S}[\rho] = \min_{\Phi \rightarrow \rho} \, \dirint{\Phi}{- \frac{1}{2} \sum \nabla_i^2}{\Phi}
\end{equation}
where $\Phi \rightarrow \rho$ indicates that the search is over all Slater determinants 
yielding the given density $\rho$. 

The deviation between the exact and the KS kinetic energy, 
$T[\rho] - T_{S}[\rho]$, is accounted for in the KS exchange-correlation functional,
\begin{equation}
\label{eq:EXC}
E_{xc}[\rho] = T[\rho] - T_{S}[\rho] + V_{ee}[\rho] -J[\rho]
\end{equation}
so that 
the original HK energy functional can be written as 
\begin{equation}
\label{eq:FHK}
F_{HK}[\rho] = T_{S}[\rho] + J[\rho] + E_{xc}[\rho]
\end{equation}
Here, $V_{ee}[\rho]$ and $J[\rho]$ denote the full and classical electronic repulsion functionals, 
respectively. Now the task of finding the minimum of the total energy functional,
\begin{equation}
\label{eq:fullenergy}
E[\rho] = F_{HK}[\rho] + \int \rho (\bs{r})v(\bs{r})\, d\bs{r} = E[\rho\{\psi_i\}]
\end{equation}
under the constraint $\int \rho (\bs{r}) d\bs{r} = N$ can be performed in orbital space 
(see Parr and Yang \cite{Parr:1989}). 
The corresponding
variational procedure yields the well-known KS equations
for the ground-state 
of a system of $N$ non-interacting electrons moving in the external potential
 \begin{equation}
\label{eq:effpot}
v^{eff}[\rho](\bs{r}) = v(\bs{r}) + J[\rho](\bs{r}) + v_{xc}[\rho](\bs{r})
\end{equation}
where 
\begin{equation}
\label{eq:vxc}
v_{xc}[\rho](\bs{r}) = \frac{\delta E_{xc}[\rho]}{\delta \rho (\bs{r})}
\end{equation}
is the KS exchange-correlation potential. More precisely, $v^{eff}[\rho](\bs{r})$ is a potential-functional, 
and Eqs.~(\ref{eq:effpot},\ref{eq:vxc}) have to be 
solved self-consistently. Self-consistency will be attained for the 
exact ground-state density, $\rho_0$, where
the KS equations with the potential $v^{eff}[\rho_0](\bs{r})$ reproduce $\rho_0$. It should be noted 
that the consistency of the KS orbitals with the LCS requirement (\ref{eq:LevyCS}) must be assured (see
Levy and Perdew~\cite{lev85:11}).

As this rigorous derivation of the KS equations shows, there is nothing that would compel a
density representation associated with $N$ non-interacting electrons: any number of non-interacting 
electrons will be permissible, even
$N=1$. Indeed, we will demonstrate in the following how 
the entire line of arguments can readily be transferred to the representation of the density in terms of 
a single (spinless) particle.
  
Obviously, any N-electron (ground-state) density function, $\rho(\bs{r})$, can be represented 
by a one-particle wave function (orbital) according to  
\begin{equation}
\label{eq:1pdensity1}
\rho (\bs{r}) = N |\phi (\bs{r})|^2
\end{equation}
where
\begin{equation}
\label{eq:1pdensity2}
\phi (\bs{r}) = (\frac{\rho (\bs{r})}{N})^{1/2}
\end{equation}
Such a representation is unique as long as $\rho(\bs{r}) \geq 0$ and $\rho(\bs{r}) \neq 0$ for finite 
values of $|\bs{r}|$. Obviously, this defines directly (that is, without invoking
the concept of non-interacting v-representability) a 1-1 mapping  
of density functions and (real) orbitals.
Next we can define a corresponding kinetic energy functional:
\begin{equation}
\label{eq:kinener}
\widetilde{T}_{S}[\rho] =  N \int \phi(\bs{r})(-\half \nabla^2)\phi (\bs{r})\, d\bs{r}
\end{equation}
Since the (real) orbital $\phi (\bs{r})$ is uniquely defined by $\rho(\bs{r})$, so is the 
kinetic energy.

It should be noted here that this definition is consistent with the Levy constrained search
procedure. The general form of an orbital reproducing the density $\rho(\bs{r})$ according to
Eq.~(\ref{eq:1pdensity1}) reads
\begin{equation}
\label{eq:genorb}
\psi(\bs{r}) = e^{ik(\bs{r})} \phi (\bs{r})
\end{equation}
where $k(\bs{r})$ is a real function. Clearly, the kinetic energy of $\psi$, 
\begin{equation}
\label{eq:phasekinener}
\left <\psi|-\half \nabla^2 |\psi \right> = \half \int (\nabla k(\bs{r}))^2 \phi (\bs{r})^2\, d\bs{r}\,\,
                              + \,\dirint{\phi}{-\half \nabla^2}{\phi}
\end{equation}
is larger than the kinetic energy of the real orbital $\phi$, 
if $k(\bs{r}) \neq const$. This means that the orbital minimizing the kinetic energy functional
for a given density is (up to a constant phase) a real function. As a consequence, 
Eq.~(\ref{eq:1pdensity2}) relates densities and orbitals, and the kinetic 
energy functional (\ref{eq:kinener}) is uniquely defined at the orbital level. We may elaborate that point 
somewhat further by considering a system where the KS orbital cannot be chosen real, e.g. in the presence
of an external magnetic field. Clearly, an orbital of the general form (\ref{eq:genorb}) is not determined
by the density alone. In addition, one has to take into account the current density, $\bs{j} = \phi^2 \nabla k$,
in order to obtain a unique definition of 
kinetic energy functional, now being a functional of both $\rho$ and $\bs{j}$, at the
orbital level. That is why a current density version of DFT must be used in the case of magnetic fields.

The next step is to introduce a correspondingly modified xc functional,
\begin{equation}
\label{eq:xcpotmod}
\widetilde{E}_{xc}[\rho] = T[\rho] - \widetilde{T}_{S}[\rho] + V_{ee}[\rho] -J[\rho]
\end{equation}
so that the functional for the total energy can be written as
\begin{equation}
\label{eq:rKSfunctional}
E[\rho] = \widetilde{T}_{S}[\rho] + J[\rho] + \widetilde{E}_{xc}[\rho] + \int \rho (\bs{r})v(\bs{r})\, d\bs{r}
\end{equation}
As in the usual KS approach, the variational search for the minimum of $E[\rho]$ under the constraint
$\int \rho (\bs{r}) d\bs{r} = N$ can equivalently be effected by a search in the space of (normalized)
orbitals $\phi(\bs{r})$. The variation of $E[\rho\{\phi\}]$ with respect to $\phi(\bs{r})$ via
\begin{equation}
\label{eq:rKSdensity} 
\rho (\bs{r}) = N |\phi(\bs{r})|^2
\end{equation}
leads to the single KS equation
\begin{equation}
\label{eq:rKS1}
\{-\half \nabla^2 + v^{eff}[\rho](\bs{r})\} \phi(\bs{r})= 
                                       \epsilon \phi(\bs{r}) 
\end{equation}
for the ground-state of a single particle moving in the effective potential
\begin{equation}
\label{eq:rKS2}
v^{eff}[\rho](\bs{r}) = v(\bs{r}) + J[\rho](\bs{r}) + \tilde{v}_{xc}[\rho](\bs{r}) 
\end{equation}
where $\tilde{v}_{xc}[\rho](\bs{r})$ is the modified xc potential deriving from $\widetilde{E}_{xc}[\rho]$, 
\begin{equation}
\label{eq:vxc1}
\tilde{v}_{xc}[\rho](\bs{r}) = \frac{\delta \widetilde{E}_{xc}[\rho]}{\delta \rho (\bs{r})}
\end{equation}
Obviously, the single GS KS orbital
has no direct physical meaning. It may be viewed as a kind of a mean orbital averaged over the 
$N/2$ spatial KS orbitals of the usual approach.

As the usual KS approach,
the radical Kohn-Sham (rKS) formulation, established by Eqs.~(\ref{eq:rKSdensity},\ref{eq:rKS1}) 
is, in principle, exact.
That is, one would obtain the exact ground-state density of the interacting
N-electron system provided the exact energy functional were available. Of course, the usual N-electron
KS formulation will be a better starting point for the use of approximative functionals, simply because
its expression for the kinetic energy, Eq.~(\ref{eq:KSKE}), will give a better approximation to the
full kinetic energy than the mean one-orbital term of Eq.~(\ref{eq:kinener}). The actual benefit 
of the rKS variant is its potential as an analytical tool, and in  
the ensuing Sec. 5 we will use that 
tool to examine the foundations of TDDFT.

Let us emphasize once again that $\tilde{v}_{xc}[\rho](\bs{r})$ is a potential-functional and
Eqs.~(\ref{eq:rKS1},\ref{eq:rKS2}) have to be solved self-consistently to yield the
exact GS density, $\rho_0 (\bs{r})$. 
As a consequence of the simple structure of the rKS equation (\ref{eq:rKS1}),
the xc potential for the exact GS density, $\rho_0 (\bs{r})$,
can be expressed according to
\begin{equation}
\label{eq:exxcpot}
\tilde{v}_{xc}[\rho_0](\bs{r}) = \frac{1}{2 \sqrt{\rho_0(\bs{r})}} \nabla^2 \sqrt{\rho_0(\bs{r})}
                                    - v(\bs{r}) - J[\rho_0](\bs{r}) + \epsilon
\end{equation}
Eq.~(\ref{eq:exxcpot}) has been used to study features of the exact KS xc potential, such as the 
asymptotic behaviour \cite{lev84:2745}. A similar equation,
arising in the ordinary KS treatment of 2-electron systems, was used to
characterize 2-electron KS xc functionals \cite{alm84:2322,bui89:4190,umr94:3827}.  

While the rKS version introduced above is the simplest possible KS-type approach,
other variants are 
conceivable in which the non-interacting system consists of $M=2$ or more electrons 
($M$ might even be larger than $N$.)
For example, in the case $M=2$ any (reasonable) density can be derived from the KS 
determinant 
\begin{equation}
\label{eq:ksdetx}
\left|\Phi^{KS}_0 \right> = |\phi_{0 \alpha} \phi_{0 \beta}|
\end{equation}
for two non-interacting spin-$\half$ particles in the spin-orbitals,
$ \phi_{0 \gamma} = \phi_0 (\bs{r}) \chi_{\gamma}(s)$, $\gamma = \alpha, \beta$, where the
spatial orbital is given by  
\begin{equation}
\label{eq:ksorb}
\phi_0 (\bs{r}) = (\frac{2 \rho_0(\bs{r})}{N})^{\frac{1}{2}}
\end{equation}
Whereas the rKS formulation is purely spatial, spin degrees-of-freedom come into play
in these $M$-electron KS variants for $M \geq 2$.

\section{Review of time-dependent DFT}
\subsection{Time dependent radical Kohn-Sham theory}
Having established the rKS formulation for the static case, we may now use
this tool to analyze td density functional theory. 

Let us assume a td external potential,
$\hat{U}(t) = \sum u(\bs{r}_i,t)$, vanishing for  $t \leq 0$, and let the system
be in its (unperturbed) ground state at $t = 0$. The solution of the td $N$-electron Schr\"{o}dinger
equation,  
\begin{equation}
\label{eq:tdSeq}
i\frac{\partial}{\partial t} \Psi(t)  = \left( \hat{H} + \hat{U}(t) \right)
                                  \Psi(t) 
\end{equation}
gives rise to an associated exact td density function, $\rho = \rho(\bs{r},t)$ 
with $\rho(\bs{r},0) = \rho_0$. As in the static case, the time development of 
the exact density can be assigned to a td orbital by generalizing Eq.~(\ref{eq:1pdensity2}): 
\begin{equation}
\label{eq:tdorb}
\phi (\bs{r},t) = (\frac{\rho (\bs{r},t)}{N})^{1/2}
\end{equation}
This is trivial. The non-trivial issue is, of course, whether one can establish a 
Schr\"{o}dinger-type equation at the single-orbital level that would allow one 
to \emph{predict} the time-development of the exact density. Because inevitably any 
(non-stationary) wave function evolving according to a td Schr\"{o}dinger equation
picks up a time- and space-dependent phase, the orbital must be written in the 
general form,  
\begin{equation}
\label{eq:tdgenorb}
\psi(\bs{r},t) = e^{ik(\bs{r},t)} \phi (\bs{r},t)
\end{equation}  
where $k(\bs{r},t)$ is a real-valued phase function, and $\phi (\bs{r},t)$ is given by 
Eq.~(\ref{eq:tdorb}).

Supposing the RG theorems valid in their original form, they will apply as well to the 
rKS formulation. Then there is a single td KS equation of the form,
\begin{equation}
\label{eq:tdKS1}
i\frac{\partial}{\partial t} \psi(\bs{r},t) = 
\{-\half \nabla^2 + v(\bs{r}) + J[\rho](\bs{r}) + u(\bs{r},t) +
\tilde{a}_{xc}[\rho](\bs{r},t)\} \psi(\bs{r},t) 
\end{equation}
governing the time evolution of $\psi(\bs{r},t)$ and, thus, of $\rho(\bs{r},t) = N |\psi(\bs{r},t)|^2$. 
Here 
\begin{equation}
\label{eq:xckernel}
\tilde{a}_{xc}[\rho](\bs{r},t) = \frac{\delta \widetilde{A}_{xc}[\rho]}{\delta \rho (\bs{r},t)}
\end{equation}
is the td xc potential associated with the rKS modification of the RG td xc functional,
$\widetilde{A}_{xc}[\rho (t)]$. 
The RG
theorems assure that such a td xc potential exists, so that, in principle, the time development
of the density can be determined exactly via Eq.~(\ref{eq:tdKS1}). In practice, of course,
one has to resort to approximations such as the widely used adiabatic local density 
approximation (ALDA). Here one uses the ordinary DFT xc potentials,
\begin{equation}
\label{eq:ALDA}
\tilde{a}_{xc}[\rho](\bs{r},t) = \tilde{v}_{xc}[\rho (\bs{r},t)]
\end{equation}
depending on time only via the time dependence of the density function, $\rho = \rho (\bs{r},t)$.
As above, the tilde indicates the rKS form of these quantities.

In the rKS version the RG theorems suggest that one can, at least in principle, 
condense the full $N$-electron td Schr\"{o}dinger equation
into a one-orbital td KS equation. Can this be true? As a step towards an answer
let us inspect how the fourth RG theorem, establishing an analogy to the KS concept 
in the time-independent DFT, works in the rKS case.

\subsection{The Runge-Gross theorems}

The KS equations have been invented as a means for determining the minimum
of the HK energy functional and thus the exact ground- state density of the interacting
$N$-electron system. In TDDFT the role of the KS equations is daringly expanded:
their time-dependend form is believed to govern, at least in principle, also the 
exact time evolution of the density of the interacting N-electron system.
The basis for that claim has been laid in a series of four theorems in a 
famous article by Runge and Gross \cite{run84:997}, in the following referred to as RG.
Let us critically review their arguments.

The first theorem (RG1) is the td analogue of the first HK theorem. It establishes
a one-to-one correspondence between td density functions, $\rho (\bs{r},t)$,
and td external potentials, $u[\rho](\bs{r},t)$, which, in turn, via the td 
Schr\"{o}dinger equation, 
\begin{equation}
\label{eq:tdSeq1}
i\frac{\partial}{\partial t} \Psi[\rho](t)  = \left( \hat{H} + \hat{U}[\rho](t) \right)
                                 \Psi[\rho](t) 
\end{equation}
determine the exact td N-electron wave functions, $\Psi[\rho](t)$ (up to a purely time-dependent
phase).

The third theorem (RG3) is the analogue to the 
second HK theorem. 
Instead of the HK energy functional, one
considers the action integral defined according to 
\begin{equation}
\label{eq:actint}
A[\rho] = \int_{t_0}^{t_1} dt\, \dirint{\Psi[\rho](t)}{i\frac{\partial}{\partial t} - \hat{H}}
                                 {\Psi[\rho](t)}
\end{equation}
We may leave any problems in this definition (see Ref.~\cite{lee01:1969}) at that
and go on further to the fourth theorem (RG4). In analogy to the ordinary KS approach, one introduces 
a kinetic-energy action functional,
\begin{equation}
\label{eq:kinact}
S_0[\rho] = \int_{t_0}^{t_1} dt \dirint{\Phi[\rho](t)}{i\frac{\partial}{\partial t} - \hat{T}}
                                 {\Phi[\rho](t)}
\end{equation}
for non-interacting particles. Here it is supposed that for a given td density function, 
$\rho (\bs{r},t)$, there exists a unique state (Slater determinant),
$ \Phi[\rho](t) $, of the non-interacting electron system. The functional
$S_0[\rho]$ is defined in analogy to the full kinetic energy action functional, 
\begin{equation}
\label{eq:fkinact}
S[\rho] = \int_{t_0}^{t_1} dt \dirint{\Psi[\rho](t)}{i \frac{\partial}{\partial t} - \hat{T}}
                             {\Psi[\rho](t)}
\end{equation}
for the original interacting electron system. As in the time-independent KS approach,
$S_0[\rho]$ replaces $S[\rho]$, the remainder, $S[\rho] - S_0[\rho]$, being transferred 
into the exchange-correlation part, $A_{xc}[\rho]$, of the full action functional (\ref{eq:actint}).
Everything seems to be completely analogous to the time-independent case.

However, there is a problem, clearly to be seen in the focus of the rKS formulation. 
Here the non-interacting state, $ \Phi[\rho](t)$, becomes a one-particle
state of the general form of Eq.~(\ref{eq:genorb}),
\begin{equation}
\nonumber
\psi[\rho](\bs{r},t) = e^{ik(\bs{r},t)} (\frac{\rho (\bs{r},t)}{N})^{1/2}
\end{equation}
so that the $S_0$ functional reads
\begin{equation}
\label{eq:rKSkinact}
\tilde{S}_0[\rho] = \int_{t_0}^{t_1} dt \int d\bs{r}
             \psi ^*(\bs{r},t) (i\frac{\partial}{\partial t} + \frac{1}{2} \nabla ^2) \psi (\bs{r},t)
\end{equation}
While the modulus of $\psi (\bs{r},t)$ is completely determined by the density $\rho (\bs{r},t)$,
the phase function $k(\bs{r},t)$ is not. Clearly, the value of $\tilde{S}_0[\rho]$  depends manifestly
on this phase function, but there is no way of determining it from the given density. This 
means that the functional $S_0$ is ill-defined at the orbital level. There are (infinitely) many orbitals
for a given density, each giving a different value for the $S_0$ functional.

Let us consider the latter argument in somewhat greater detail.
Inserting the form (\ref{eq:tdgenorb}) of the orbital in the 
integrand of the $S_0[\rho]$ functional one readily obtains 
\begin{equation}
\label{eq:s0}
\dirint{\psi}{i \frac{\partial}{\partial t} + \half \nabla ^2  }{\psi} = 
\dirint{\phi}{ \half \nabla ^2 }{\phi} - \dirint{\phi}{(\nabla k)^2 }{\phi} - \dirint{\phi}{\dot{k}}{\phi}
\end{equation}
This means that besides the density here also the gradient of the phase function,
$\nabla k(\bs{r},t)$, and the time derivative,
$\dot{k}(\bs{r},t)$, is needed. Indeed, the latter information can 
be derived from the density to a certain, yet insufficient extent.  
Obviously, the orbital not only is to reproduce the density, but also
to fulfill a td Schr\"{o}dinger equation (SE) of the form
\begin{equation}
\label{eq:tdKS2}
i\frac{\partial}{\partial t} \psi(\bs{r},t) = 
\{-\half \nabla^2 + w(\bs{r},t)\} \psi(\bs{r},t) 
\end{equation} 
where $w(\bs{r},t)$ is a local td potential yet to be determined.
Therefore, the continuity equation
\begin{equation}
\label{eq:conteq}
\frac{d}{dt}\phi^2 + \nabla \bs{j} = 0
\end{equation}
applies to the orbital, where the current density is given by
\begin{equation}
\label{eq:cdens}
\bs{j} = \phi^2 \nabla k
\end{equation}
As a consequence, it is possible to determine 
$\nabla k$ from $\rho$ and $\dot{\rho}$, respectively. 
(A mathematical complication may arise here due to the 
requirement that $\nabla \times (\bs{j}/ \phi^2)$ must vanish.)
Further, if $\nabla k$ is given 
(at any time), then also $k(\bs{r},t)$ is determined, though only up to a pure
time-dependent function, $\alpha(t)$. But as Eq.~(\ref{eq:s0}) clearly shows, the latter indefiniteness
of the phase function prevents the $S_0$ functional to become well-defined. The time integral 
$\int \dot{\alpha}(t) dt$ on the r.h.s. of Eq.~(\ref{eq:s0}) leads to a completely undetermined constant in the 
definition of the $S_0$ functional. Note that this does not mean just a uniform shift of the
$S_0$ values, which, of course, would drop out in a variational search for stationary points.

So far we have not specified the local potential $w(\bs{r},t)$ in Eq.~(\ref{eq:tdKS2}), but only
assumed that such a potential exists, e.g., as a consequence of  
the first RG theorem (RGI) applied to the non-interacting
KS system (of one electron). But in the rKS version, the RGI result can be obtained in a direct way 
(thereby proving the one-orbital td v-representability of any ``reasonable'' density), allowing us
even to give 
an explicit expression for  $w(\bs{r},t)$. This is achieved
by inserting the 
ansatz (\ref{eq:tdgenorb}) in the 
SE (\ref{eq:tdKS2}). Separating the real and imaginary parts yields the following two equations: 
\begin{equation}
\label{eq:Re}
w(\bs{r},t) = \frac{\nabla^2 \phi}{2 \phi} - \half (\nabla k)^2 - \dot{k}  
\end{equation}
and
\begin{equation}
\label{eq:Im}
\dot{\phi} + \nabla k \nabla \phi + \half \nabla^2 k \,\phi = 0
\end{equation}
Obviously, the latter equation reproduces the continuity equation 
(\ref{eq:conteq}), whereas  
the former gives an explicit expression for the local td potential, $w(\bs{r},t)$.
Since both $\nabla k$ and $k$ result from $\rho$ as discussed above, $w(\bs{r},t)$ is
determined by $\rho$  up to a  purely time-dependent function, namely $\dot{\alpha}(t)$.
This shows that even the explicit form of the potential
 is of no avail to determine $\alpha(t)$. 
Even if the value of  $\alpha(t)$, was given (or fixed) at an initial time, it cannot be determined for 
later times by solving the td SE (\ref{eq:tdKS2}) due to the corresponding indefiniteness of $w(\bs{r},t)$.

Thus, the rKS formulation inevitably points our view on the problem of the undetermined
purely time-dependent phase functions corrupting 
the RG action integral functionals. The phase problem arises not only in the functionals
of the non-interacting KS system but already in the functional (\ref{eq:actint}) for  
the original interacting $N$-electron system. 
When one consults the RG paper~\cite{run84:997}
with regard to this issue, one finds that the phase problem, being discussed in the beginning of the paper,
gets lost in the matrix element
$\dirint{\Phi(t)}{i \frac{\partial}{\partial t} - \hat{T} - \hat{W} - \hat{V}(t)}{\Phi(t)}$
after Eq.~(11). Here $\hat{V}(t)$ is  the external td potential of the physical
system under consideration. This potential does not contain a function $C(t)$ that would
cancel the time derivative of the phase function in $\Phi(t)$.
Erroneously, RG argue here as if this potential was the potential 
$\tilde{V}(t)$ according to the RGI theorem, that is, the potential invoked in the td SE for the 
$N$-electron wave function, $\Phi(t)$, corresponding to the considered density.
(Later TDDFT papers and virtually all review articles inconspicuously leap over the phase problem in
the action integral functionals.)

Whereas there is still wide-spread confidence in the RG foundations of TDDFT, their breakdown due 
to the phase problem has been clearly analyzed and expressed by R. van Leeuwen 
already several years ago \cite{lee98:1280,lee01:1969}. In his 2001 review article~\cite{lee01:1969} he draws 
the following conclusion: ``\emph{We therefore conclude that time-dependent density-functional 
theory can not be based on the usual variational principle, and indeed attempts to do so have 
led to paradoxes.}'' Let us note that besides the phase problem, van Leeuwen
also analyzed correctly the non-stationarity of the RG action integral functionals, exposing
another fault line in the original RG argumentation. But why have van Leeuwen's revelations not triggered
stronger shock waves in the TDDFT community and beyond? Apparently because by the time of van Leeuwen's 
analysis the leading actors in the field had come to the conclusion that the KS equations could be
established directly without the necessity of resorting to a variational 
principle~\cite{gross:1995,goe98:265}. Sharing that conviction, van Leeuwen 
communicated the reassuring message that TDDF is valid, though only in a new shape featuring the so-called 
Keldysh Green's function technique~\cite{kel65:1018}.  
In the ensuing subsection 5.C, we will have a closer look at the non-variational KS theory.

\subsection{Kohn-Sham equations without a variational principle?}

In the derivation of the static KS equations three elements are essential: \emph{i}) a universal
energy functional (HKI); \emph{ii}) a variational principle for the exact ground-state density (HKII); 
and \emph{iii})
a functional for the kinetic energy of non-interacting electrons defined at the orbital level (KS).
Runge and Gross have pursued a strictly analogous approach in order to establish a basis for TDDFT. As first 
analyzed by van Leeuwen and corroborated here,
this endeavor must be viewed as foundered in each of the three essentials. 

But is there  
a different route to establishing td KS equations? Within the TDDFT community,   
the generally accepted view is that this is the case. Indeed,
the first RG theorem offers a shortcut to KS-type equations.
Applying RGI to the case of non-interacting $N$ electrons, one can establish the mapping
\begin{equation}
\label{eq:map}
\rho(\bs{r},t) \longrightarrow w[\rho](\bs{r},t)  
\end{equation}
so that the td KS-type equations
\begin{equation}
\label{eq:tdKS3}
i\frac{\partial}{\partial t} \psi_j(\bs{r},t) = 
\{-\half \nabla^2 + w[\rho](\bs{r},t)\} \psi_j(\bs{r},t),\;\;j = 1,\dots,N  
\end{equation} 
allow one to calculate the density $\rho(\bs{r},t)$ from the orbitals $ \psi_j(\bs{r},t)$.
The KS potential in Eq.~(\ref{eq:tdKS3}) can be written in a more familiar form,
\begin{equation}
\label{eq:tdKS4}
w[\rho](\bs{r},t) = v_{ext}(\bs{r},t) + J[\rho](\bs{r},t) + v_{xc}[\rho](\bs{r},t) 
\end{equation}
where  $v_{ext}(\bs{r},t) = v(\bs{r}) + u(\bs{r},t)$ comprises
the static and td external potentials of the system under consideration. 
Apparently, Eq.~(\ref{eq:tdKS4}) serves as a definition of an
xc potential-functional $v_{xc}[\rho](\bs{r},t)$ by subtracting two known potentials
from the unknown KS potential-functional $w[\rho](\bs{r},t)$ 
(see Refs.~\cite{hei02:9624,mar04:427,bur05:062206}). 
At least formally, everything looks as one
would expect. 
But is this really the solution? Once more, the rKS formulation allows for a closer
inspection of what we have got, because here the potential $w[\rho](\bs{r},t)$ 
can be given in an explicit form.

Indeed, as the analysis of Sec.~5.B has shown, for a density $\rho(\bs{r},t)$ there  exists a
single-particle td SE (Eq.~\ref{eq:tdKS2}),    
\begin{equation}
\label{eq:tdKS2x}
i\frac{\partial}{\partial t} \psi(\bs{r},t) = 
\{-\half \nabla^2 + w[\rho](\bs{r},t)\} \psi(\bs{r},t) 
\end{equation} 
with the td local potential (Eq.~\ref{eq:Re}),
\begin{equation}
\label{eq:Rex}
w[\rho](\bs{r},t) = \frac{\nabla^2 \phi}{2 \phi} - \half (\nabla k)^2  - \dot{k}  
\end{equation}
which is determined by the density up to a purely td
function ($\dot{\alpha}(t)$). Let us note that now the
indefinite td function is no longer relevant, 
because it does not affect the resulting density.
But it seems that Eqs.~(\ref{eq:tdKS2x},\ref{eq:Rex}) lack any predictive power. 
They hold for any density, and one may wonder 
how the time-development of
the exact density of the interacting $N$-electron system,
$\rho_0(\bs{r},t)$, could be determined unless $\rho_0$
is already known and used to construct $w[\rho_0](\bs{r},t)$.

At this point it is instructive to inspect the
more transparent case of static DFT. 
Let us assume for a moment that there is no second 
Hohenberg-Kohn theorem (HKII) and, thus, no variational principle. As above, however,
one has a shortcut to KS-type equations (now applying the HKI theorem
to the non-interacting system). In the rKS variant, the corresponding single KS-type equation  
can explicitely be constructed
(by inserting the ansatz (\ref{eq:1pdensity2}) 
in the one-particle Schr\"{o}dinger equation):   
\begin{equation}
\label{eq:mKS1}
\{-\half \nabla^2 + w[\rho](\bs{r})\} \phi(\bs{r}) = \epsilon  \phi(\bs{r})
\end{equation}      
Here the potential  
\begin{equation}
\label{eq:mKS2}
w[\rho](\bs{r}) = \frac{\nabla^2 \sqrt{\rho}}{2 \sqrt{\rho}} + c
\end{equation}
is determined by the density $\rho(\bs{r}) = N \phi(\bs{r})^2$ (up to a constant $c$).
Eqs.~(\ref{eq:mKS1},\ref{eq:mKS2}) show that any (reasonable) density is non-interacting 
(one-electron) v-representable. 
But, clearly, the potential-functional $w[\rho](\bs{r})$ of Eq.~(\ref{eq:mKS2}) as such
is of no avail for determining the exact ground-state density, $\rho_0$. According to
the successive steps,
\begin{equation}
\nonumber 
\rho(\bs{r}) \longrightarrow w[\rho](\bs{r}) \longrightarrow \mathrm{(Eq.\phantom{,}\ref{eq:mKS1})} 
\longrightarrow \phi(\bs{r}) \longrightarrow \rho(\bs{r})
\end{equation}
any density $\rho(\bs{r})$ will only reproduce itself. Obviously, 
the potential-functional (\ref{eq:mKS2}) is trivial, i.e. without
physical meaning.
In the variationally derived  KS equation (\ref{eq:rKS1}), by contrast, the potential-functional 
$v^{eff}[\rho](\bs{r}) = v(\bs{r}) + J[\rho](\bs{r}) + \tilde{v}_{xc}[\rho](\bs{r})$ according 
to Eq.~(\ref{eq:rKS2}) is of completely different type. The density will change 
in the course of the iterative solution of the KS equation and will (eventually) converge to
the exact (or approximate) ground-state density $\rho_0$. Only for $\rho_0$, the KS equation 
(with $v^{eff}[\rho_0](\bs{r})$) will reproduce the initial density $\rho_0$. At this point,
the non-trivial and the trivial potential become identical (up to a constant), 
$v^{eff}[\rho_0](\bs{r}) = w[\rho_0](\bs{r}) + c$, as can be seen by comparing Eqs.~(\ref{eq:exxcpot})
and (\ref{eq:mKS2}). 

Yet this is not the end of the story.
While hardly discussed in the literature, the triviality of the KS potential-functionals arising in the 
HKI/RGI shortcut is well-known among the DFT theoreticians. Rather than ``defining''
an xc potential-functional by subtracting the given one-particle potential, $v(\bs{r})$, and
the Hartree potential, $J[\rho](\bs{r})$, from the trivial KS potential-functional, $w[\rho](\bs{r})$, a
non-trivial xc potential-functional can be established by the following partioning~\cite{lee01:1969,gro06:pc}:
\begin{equation}
\label{eq:ntriv}
w[\rho](\bs{r}) = v_{ext}[\rho](\bs{r}) + J[\rho](\bs{r}) + v_{xc}[\rho](\bs{r})
\end{equation}
Here $v_{ext}[\rho](\bs{r})$ is the potential-functional established by the HKI theorem for the
interacting $N$-electron system, yielding  
\begin{equation}
\label{eq:ntrivx}
v(\bs{r}) = v_{ext}[\rho_0](\bs{r})
\end{equation}
for the exact ground-state density, $\rho_0$. Letting aside the $v$-representability problem
in the $v_{ext}[\rho]$ potential-functional, one can easily see that the xc potential-functional
defined by Eq.~(\ref{eq:ntriv}) is the same as that arising in the variational derivation. 
Having established the existence of a non-trivial xc potential-functional, one may now use it
in the KS equation in the familiar way, that is, together with the given one-particle potential
$v(\bs{r})$ of the considered system. What one gets is fully equivalent to the variationally derived
result (Eqs.~\ref{eq:rKS1},\ref{eq:rKS2}): 
a KS equation that does not reproduce any density except for the exact ground-state
density $\rho_0$, thereby offering the possibility of determining $\rho_0$ by a self-consistency
procedure. The only difference between the variational and non-variational derivation is that
in the former case the iterative procedure represents a well-defined search for a minimum 
on an energy surface, whereas in the non-variational approach the final step 
amounts to an \emph{ad hoc} ansatz of a \emph{regula falsi} type, for which the question of
convergence remains open. 
In practice, of course, this distinction does not matter, because, wittingly or unwittingly,
the variational KS derivation will serve as
a safeguard for the convergence of the self-consistency cycle in the non-variational approach.
 
Now we may come back to the time-dependent case. Lacking variationally derived KS equations,
one can, nevertheless, establish a ``non-trivial'' xc potential-functional in analogy 
to Eq.~(\ref{eq:ntriv}):
\begin{equation}
\label{eq:ntrivtd}
w[\rho(t)](\bs{r},t) = v_{ext}[\rho(t)](\bs{r},t) + J[\rho(t)](\bs{r},t) + v_{xc}[\rho(t)](\bs{r},t)
\end{equation} 
where $v_{ext}[\rho(t)](\bs{r},t)$ denotes the potential-functional established via the 
RGI theorem (analogue to HKI). For the exact density trajectory, $\rho_0(\bs{r},t)$, of the 
interacting $N$-electron system with the external potential $v_{ext}(\bs{r},t)$ the RGI 
potential-functional gives 
\begin{equation}
\label{eq:ntrivtdx}
v_{ext}[\rho_0(t)](\bs{r},t) = v_{ext}(\bs{r},t)
\end{equation}
Again, we may use the non-trivial xc potential-functional $v_{xc}[\rho(t)](\bs{r},t)$
in the (radical) KS equation together with  $v_{ext}(\bs{r},t)$, giving rise to
the td one-orbital Schr\"{o}dinger equation
\begin{equation}
\label{eq:tdKSx}
i\frac{\partial}{\partial t} \psi(\bs{r},t) = 
\{-\half \nabla^2 + v_{ext}(\bs{r},t) + J[\rho(t)](\bs{r},t) + v_{xc}[\rho(t)](\bs{r},t)\} \psi(\bs{r},t)
\end{equation} 
which is ``correct'' only for the exact density $\rho_0(\bs{r},t)$. In the latter case 
the effective potential in Eq.~(\ref{eq:tdKSx}) becomes $w[\rho_0](\bs{r},t)$, that is, the trivial
KS potential-functional of Eq.~(\ref{eq:Rex}) taken at the exact density $\rho_0(\bs{r},t)$.
This means we have returned to our starting point and the original 
question: 
Will the KS Eq.~(\ref{eq:tdKSx}) with the potential $w[\rho_0](\bs{r},t)$ allow us to
determine the time evolution of the exact density $\rho_0(\bs{r},t)$ without knowing it beforehand?

Let us first note that there is no longer a self-consistency cycle providing for any
feedback to the exact solution. Eq.~(\ref{eq:tdKSx}) would have to be solved by time-propagation
all the way ``along'' the exact density trajectory $\rho_0(t)$. 
In practice, that would mean to propagate an ``incorrect'' equation, using a guessed xc potential,
and starting from an approximate ground state density at, say, $t=0$. But apart from
such practical reservations, the basic question is whether the propagation would succeed
given the exact xc potential-functional and the exact density at $t=0$. 
But even in this ideal 
case the answer is seen to be negative. The problem here is that the mapping 
$\rho(t) \rightarrow w[\rho(t)]$ (Eq.~\ref{eq:map}) is ``non-local'' in time (or non-instantaneous).
What does this mean? An
instantaneous potential-functional, for example, is the Hartree potential, $J[\rho(t)](\bs{r},t)$:
the density at given time $t$ determines the Hartree potential at the same moment $t$. But
the situation is not as simple in the case of the KS potential-functional. 
This can be seen by inspecting again the KS
potential-functional~(\ref{eq:Rex}) of the rKS formulation. The first contribution on the 
r.h.s. of Eq.~(\ref{eq:Rex}) is instantaneous; the second term, depending on 
the gradient of the phase function $k(\bs{r},t)$, requires the first time-derivative
of the density according to the discussion in Sec.~V.B; and the third term, 
being the time-derivative of the phase function $k(\bs{r},t)$, can only be determined if the
second time-derivative of the density is available. A similar temporal non-locality must 
be expected for $v_{xc}[\rho(t)]$ and $v_{ext}[\rho(t)]$. The consequence for the 
time-propagation of Eq.~(\ref{eq:tdKSx}) is obvious: the second time-derivative of the density 
is not determined by the development through a given time $t$ (``past'') so that the
potential at time $t$ is undefined unless one takes into account also the density trajectory
beyond the point $t$ (``future''). The inevitable conclusion is that 
the non-variational td KS equations 
are not suitable for \emph{predicting} the time-development 
of the exact density. They would allow one to \emph{reproduce} the time-development
of the density at the orbital level, provided the td density is already given, e.g., from a 
solution of the full $N$-electron td Schr\"odinger equation, which is of course
without practical use. 
At a purely formal level of the theory, this crucial difference between 
predicting and merely reproducing is well concealed and, thus, easy to overlook.

In view of this finding,
the ``causality problem'' in the TDDFT linear response equations is to be seen from a
new perspective. Rather than being the consequence of improperly defined xc functionals, as 
van Leeuwen and others have supposed, the problem seems to reflect
the basic inadequacy of a linear response treatment for the unphysical, merely formal td
KS equations.

\subsection{Linear response in the adiabatic approximation}

So far we have used the rKS concept as a tool to analyze some basic aspects of the TDDFT approach.
Let us finally take a view at the structure of the results to be expected at the linear 
response (LR) level of the theory.

Using the adiabatic approximation for the td xc potential, both in the usual and radical
KS versions, leads to the RPA-like equations given by Eq.~(\ref{eq:mateqs}), where the blocks
of the secular matrix,
$\bs{A}$ and $\bs{B}$, are constant ($\omega$-independent) matrices. The excitation energies,
$\omega_m = E_m - E_0$, are obtained as the eigenvalues of the pseudo-eigenvalue problem
\begin{equation}
\label{eq:RPAeig}
\left( \begin{array}{cc}
                      \bs{A}   &  \bs{B}\\
                      \bs{B}^* & \bs{A}^*
         \end{array} 
         \right)
                     \left( \begin{array}{c}
                                  \ul{x}_m\\
                                  \ul{y}_m
                                  \end{array}
                            \right) = 
                         \omega_m \left( \begin{array}{c}
                                  \ul{x}_m\\
                                  - \ul{y}_m
                                  \end{array}
                            \right)
\end{equation} 
The transition moment associated with the $0 \rightarrow m$ transition derives from 
the corresponding (specifically normalized) pseudo-eigenvector components according to 
\begin{equation}
\label{eq:RPAtranmom}
\dirint{\Psi_m}{\hat{D}}{\Psi_0} = \sum_{a,k} \left (x_{ak,m}^* d_{ak} + y_{ka,m}^* d_{ka}\right) 
\end{equation}
The manifold of excitations obtained from these equations is determined by the 
configuration space of the secular matrix block $\bs{A}$ (note that the RPA pseudo-eigenvalues
occur in pairs having positive and negative values, respectively). For the ordinary KS approach
this means that the excitation manifold is that 
of the $p$-$h$ or single excitations (with respect to the GS KS determinant). Here
each spatial $p$-$h$ configuration gives rise to 4 (primitive) spin states, from which
one singlet and three (degenerate) triplet states can be formed. Let $n_o = N/2$ and $n_v$ 
denote the number of occupied and virtual
spatial KS orbital. Then the KS LR excitation manifold comprises $4 n_o n_v$ solutions. The 
full excitation manifold of $N$ interacting electrons is, of course, much larger, because
double and higher excitations come into play. It is thought that the restriction to 
single excitations is a consequence of the adiabatic approximation and the missing 
double and higher excitations would be accounted for by going beyond that approximation. Supposing
that the exact td xc potential, $a_{xc}[\rho](\bs{r},t)$, exists, one would arrive at the same type of
equations as in Eq.~(\ref{eq:mateqs}), but now with $\omega$-dependent matrices, 
$\bs{A}(\omega)$ and $\bs{B}(\omega)$. In principle, this could lead to an  enhanced
excitation manifold. 

Let us now inspect the excitation manifold in the rKS case. The LR equations within 
the adiabatic approximation have the 
same structure (Eq.~\ref{eq:mateqs}) as those of the usual KS approach, but there is only one
occupied spatial KS orbital ($n_o =1$). As a consequence, the excitation manifold comprises only
$n_v$ excitations, that is, the excitations out of a single (average) KS orbital. Moreover,
any spin degrees-of-freedom are missing, and, even if one
assigns the $n_v$ spatial excitations to singlets, any triplet excitations are absent. 
We have argued that the td treatment in the rKS framework is as legitimate or not legitimate as the usual
KS approach. This means that here the hypothetical non-adiabatic td xc potential, 
$\tilde{a}_{xc}[\rho](\bs{r},t)$, must not only account for the double and higher excitations but 
has to restore already the single excitation manifold. A generation of triplet excitations appears to be
completely impossible, because the density function $\rho(\bs{r},t)$ and, thus, 
$\tilde{a}_{xc}[\rho](\bs{r},t)$, does not bear information on the spin degrees-of-freedom. 
Let us note that the absence of triplets in the rKS version does not constitute an 
inconsistency with the $N$-particle KS case. In principle, the standard KS approach too does not allow for 
triplets, because the perturbing td potential has to be local, and a local potential cannot excite
triplets from a singlet ground state. This fact is often repressed because, in a technical sense, the
usual TDDFT LR equations do yield triplet excitations (albeit with vanishing intensities). In the rKS
scheme triplet excitations are neither accessible basically nor technically.

As was noted in Sec.~4, the rKS variant is only the limiting case of more general $M$-electron KS
schemes, where the number $M$ of non-interacting electrons may even exceed $N$. In the latter case 
the adiabatic approximation would produce more single excitations than the original interacting $N$-electron
system, which would mean that the non-adiabatic td xc potentials must eliminate spurious solutions 
introduced at the adiabatic level of theory.  

This shows that the well-known excitation manifold or ``counting'' problem of the
LR form of TDDFT is further aggravated in the rKS (and $M$-electron KS) variants. Whereas it
cannot be excluded that a hypothetical energy dependent xc potential beyond the 
adiabatic approximation might restore the single-excitation manifold of the ordinary KS scheme and,
moreover, generate double and higher excitations, it appears more convincing to see this problem
as an indication of the invalidity of the TDDFT equations.  

\section{Summary and conclusions}

The title of this paper poses the question whether TDDFT is formally
exact. What have we learned to answer that question?
Let us summarize the three main topics of our investigation.

First, we have observed that an error is introduced
both in the td and static KS linear response theory if 
the perturbing (external) potential is given by a non-local operator. 
This error, resulting from the neglect of the $h$-$h$ and $p$-$p$ matrix elements 
of the perturbing operator in the KS response equations, is of second order in the 
electronic repulsion. Yet for a local (multiplicative) potential 
no logical inconsistency arises, because the absence of $h$-$h$ and $p$-$p$ matrix elements 
does not imply a loss of information of the local operator. As stated in Theorem 1 of
Sec. 3, this is a remarkable general property of local operators. It is a well-known
consistency requirement of DFT that the external potentials must be local. One can also
extend the HK and KS approach to general non-local potentials, but that would require
the incorporation of the non-local potentials already in the definition of the 
HK functional, $F_{HK}[\rho]$, and, accordingly, in the KS xc potential, $v_{xc}[\rho]$. As a 
consequence, the functionals would no longer be universal but depend on the respective 
non-local potentials. 
  
The problem of the non-local operators reminds us
that the KS LR equations cannot be viewed as having an unconditional physical meaning. In fact,
their validity derives from the underlying theory. 
The CPKS equations, for instance, are founded on a valid first-order perturbation theory
for the KS equations for the perturbed $N$-electron system. In the td case the validity of the response
equations would presuppose that the time-dependent extension of the KS equations be correct, 
that is, the td KS equations establish a formally exact approach to the time-development 
of the $N$-electron density.

Secondly, we have discussed an utmost simplification of the KS concept,
referred to as radical Kohn-Sham (rKS) approach. Here the ground-state density is
obtained from a single one-particle KS equation supposing a correspondingly modified xc potential.
In principle, the rKS form of the theory is as legitimate as the usual N-particle KS approach.
Whereas the ordinary KS approach will certainly be better 
suited for developing practical computational schemes, it is not inconceivable that the rKS variant 
will have some computational potential as well. More importantly though, 
the rKS formulation represents a useful pedagogical and  
analytical tool, and as such it has been used here
to elucidate basic aspects of DFT and TDDFT. 
 
In the td extension of the rKS approach, having the same validity status as the ordinary td KS
theory, a single td one-particle KS equation, though involving a possibly very 
complicated td xc potential,
would allow us to determine exactly the time-development of the density 
function of the full interacting $N$-electron system, thus bypassing the $N$-electron  
time-dependent Schr\"{o}dinger equation.
Given the richness of the phase relations of the full N-electron wave function, the 
spin degrees-of-freedom,
even the permutation symmetry, all the wealth of information seemingly absent in a 
one-particle orbital or
the one-particle density function, 
the possibility of predicting the exact time-development at a fictitious one-particle level must appear
fantastic. But this is what TDDFT implies. Is that expectation justified? 
Guided by the rKS approach, in the third step, we have critically reexamined the RG foundation of
TDDFT. What has become apparent here is a phase problem corrupting the definitions of 
the RG action integral functionals, as already recognized and analyzed 
by van Leeuwen in a different way~\cite{lee01:1969}. Our findings fully confirm van Leeuwen's conclusion
that the RG foundation of TDDFT, based on analogues to the HKI, HKII, and KS theorems, 
is invalid. 

But there is an alternative way of establishing the KS equations of TDDFT
without invoking a variational principle, as formulated by Gross, van Leeuwen and others. 
Again, this idea can be analyzed at the rKS level. 
The mere existence of a one-particle KS type equation \emph{reproducing} the exact 
td density is almost a triviality within the rKS framework. But due to temporal
non-locality the corresponding KS potential-functional does not allow 
 one to \emph{predict} the time-development of the exact density of the interacting
$N$-electron system under consideration (without introducing the information of the exact density
at some point). In its present shape, TDDFT is just a formal 
theory without any predictive power. 
As we have argued in Sec. 5.C, 
van Leeuwen's 
construction of Keldysh functionals does not cure that basic deficiency.

Given the original RG foundation of TDDFT invalid and the design of a KS theory without a variational 
principle an illusion, it appears hardly possible to escape the conclusion that the idea of TDDFT,
that is, the idea of a formally exact method for predicting the time-development of an interacting $N$-electron
system at the orbital level, must be abandoned.
The TDDFT approach (in linear response form) was first 
introduced 25 years ago 
as an analogue to TDHF (or RPA) before any attempts at a rigorous foundation had been made.   
Without the RG theorems or another viable justification, the theory would be set back to the 
status it had in its beginning: 
an empirically ``corrected'' version of the RPA~\cite{cai02:5543}. 
While TDDFT (LR) may afford an improvement
over the RPA description, it cannot escape the RPA limitation of being an approximate method for 
singly excited states.

\section*{Acknowledgements}

We are indebted to Martin Head-Gordon for giving us the hint that 
the loss of the $h$-$h$ and $p$-$p$ matrix elements of a local external potential
operators in the Kohn-Sham linear response equations does not constitute a 
logical inconsistency of DFT. We would like to thank
H.-D. Meyer, M. Lein, J. Brand, P. Pl\"{o}ck, and K. Sch\"{o}nhammer for clarifying discussions.
Moreover, we owe valuable insights  
to E.K.U. Gross, A. G\"{o}rling, and F. Furche, who kindly were open to 
discussing our controversial points of view.   

A. D. gratefully acknowledges financial support by the Deutsche Forschungsgemeinschaft
as an ``Emmy Noether'' fellow.

\newpage
\appendix
\renewcommand{\theequation}{A.\arabic{equation}}
\setcounter{equation}{0}
\section*{Appendix: Matrix representation of local operators}
\noindent
For matrix representations of local one-particle operators the following
uniqueness theorem holds:\\
\emph{Theorem}: If two local operators, $v(\bs{r})$ and $w(\bs{r})$, have the same $p$-$h$
matrix elements with respect to an arbitrary partitioning of a (complete) one-particle
basis into hole($h$) and particle($p$) states, they can differ only by a constant
$\lambda$, that is, $ w(\bs{r}) =  v(\bs{r}) + \lambda$.\\
\emph{Proof}: Let $\phi_s(\bs{r}), \, s = 1, 2, \dots ,$ denote the functions (orbitals) of a 
one-particle basis
and assume a partitioning of the orbitals such that $\phi_1(\bs{r}), \dots,\phi_n(\bs{r})$
are referred to as occupied or hole states and $\phi_{n+1}(\bs{r}), \phi_{n+2}(\bs{r}), \dots$
as unoccupied or particle states. Consider two local operators $v(\bs{r})$ and $w(\bs{r})$ having
the same $p$-$h$ matrix elements,
\begin{equation}
\label{eq:app1}
\dirint{\phi_a}{v}{\phi_k} =\dirint{\phi_a}{w}{\phi_k},  \,\, k \leq n,\, a > n 
\end{equation}
This means that the $p$-$h$ matrix elements of the difference operator, 
$\lambda(\bs{r}) = w(\bs{r}) -v(\bs{r})$ vanish:
\begin{equation}
\label{eq:app2}
\lambda_{ak} = \dirint{\phi_a}{w - v}{\phi_k} = 0, \,\, k \leq n, \, a > n 
\end{equation}
Now consider the $n$ functions $\lambda(\bs{r}) \phi_l(\bs{r}),\, l \leq n$.  
These functions may be expanded in terms of the basis functions,
\begin{eqnarray}
\lambda(\bs{r}) \phi_l(\bs{r}) & = & \sum^{\infty}_{s=1} \lambda_{sl} \,\phi_s(\bs{r})\\ 
                                & = & \sum^n_{k=1} \lambda_{kl} \,\phi_k(\bs{r}), \,\,\, l \leq n
\end{eqnarray}
yielding finite linear combinations 
as a consequence of Eq.~(\ref{eq:app2}). Obviously, the latter equations can be brought to
diagonal form by a suitable unitary transformation:
\begin{equation}
\lambda(\bs{r}) \widetilde{\phi}_k(\bs{r}) = \tilde{\lambda}_{kk}\widetilde{\phi}_k(\bs{r}), \,\,k \leq n
\end{equation}
This means that
all transformed diagonal matrix elements must be equal,
$\widetilde{\lambda}_{kk} = \lambda, \, k \leq n$, and the difference potential is 
constant: $\lambda(\bs{r}) \equiv \lambda$.

The proof given here shows that the theorem can also stated as follows: Any local operator with 
vanishing $p$-$h$ matrix elements with respect to a complete basis set and an arbitrary partitioning
into $p$ and $h$ states must be a constant. An apparent objection is: What about a diagonal representation
of the operator? The answer is that local operators cannot be diagonalized properly, that is, 
in the Hilbert space of $l^2$ functions.

An interesting question arising in this context is if it is possible to reconstruct a local
operator $v(\bs{r})$ (up to a constant), if only its $p$-$h$ matrix elements are given. It seems that this is 
not possible except for the special case $n=1$ (one occupied state). Let us first inspect the case
$n = 1$ and let $\phi_1(\bs{r})$ be the single $h$ orbital. Expanding $v(\bs{r})\phi_1(\bs{r})$ yields
\begin{equation}
v(\bs{r})\phi_1(\bs{r}) = v_{11} \phi_1(\bs{r}) + \sum^{\infty}_{a=2} v_{a1} \phi_a(\bs{r}) 
\end{equation}
where $v_{pq} = \dirint{\phi_p}{v}{\phi_q}$ denote the matrix elements of $v(\bs{r})$. 
Dividing this expression by $\phi_1(\bs{r})$ yields an explicit representation 
\begin{equation}
v(\bs{r}) = v_{11}  + \sum^{\infty}_{a=2} v_{a1} \phi_a(\bs{r}) \phi_1(\bs{r})^{-1} 
\end{equation} 
which reconstructs $v(\bs{r})$ in terms of the $p$-$h$ matrix elements $v_{a1}$ up to a constant,
being here the (single) $h$-$h$ matrix element, $v_{11}$. 

In obvious generalization of the case $n = 1$ one may proceed as follows. Let there be
$n$ occupied orbitals, $\phi_1(\bs{r}), \dots , \phi_n(\bs{r}), n > 1$. Expanding the products,
$v(\bs{r})\phi_i(\bs{r})$, gives
\begin{equation}
v(\bs{r})\phi_i(\bs{r}) = \sum^n_{k=1} v_{ki} \phi_k(\bs{r})  + \sum^{\infty}_{a=n+1} v_{ai} \phi_a(\bs{r}),
                                                               \,\,\, i \leq n 
\end{equation}
where the summation on the rhs has been split into  $h$ and $p$ parts. As in the proof above, the
$h$-$h$ block of the $v$ matrix can be diagonalized by a suitable unitary transformation, yielding
\begin{equation}
v(\bs{r})\widetilde{\phi}_i(\bs{r}) = \widetilde{v}_{ii} \widetilde{\phi}_i(\bs{r})   
                     + \sum^{\infty}_{a=n+1} \widetilde{v}_{ai} \phi_a(\bs{r}), \,\,\, i \leq n 
\end{equation}    
Here $\widetilde{v}_{ai}$ denote the transformed $p$-$h$ matrix elements of $v(\bs{r})$. Dividing
these equations by the respective transformed occupied orbital, $\widetilde{\phi}_i(\bs{r})$, leads to
$n$ different representations of $v(\bs{r})$,
\begin{equation}
v(\bs{r}) = \widetilde{v}_{ii}   
         + \sum^{\infty}_{a=n+1} \widetilde{v}_{ai} \phi_a(\bs{r})\widetilde{\phi}_i(\bs{r})^{-1} , \,\,\, i \leq n 
\end{equation} 
However, this does not solve the problem because the transformed $p$-$h$ matrix elements, $\widetilde{v}_{ai}$
cannot be determined without diagonalization of the $h$-$h$ block of the $v$ matrix, that is,
without the knowledge of the $h$-$h$ matrix elements. Thus, it appears that one encounters a similar
situation as in the theoretical foundation of DFT, where the Hohenberg-Kohn theorem states 
the existence of an universial xc functional without any constructive means.

\end{document}